\documentclass[
aps,floatfix,nofootinbib]{revtex4}
\usepackage{latexsym}
\usepackage{mathptm}
\usepackage{amsmath}
\usepackage{amssymb}
\usepackage{graphicx}
\def\beq{\begin{equation}}
\def\eeq{\end{equation}}
\def\beqn{\begin{eqnarray}}
\def\eeqn{\end{eqnarray}}
\newcommand{\f}{\begin{equation}}
\newcommand{\ff}{\end{equation}}
\newcommand{\be}{\begin{equation}}
\newcommand{\ee}{\end{equation}}
\newcommand{\barray}{\begin{array}}
\newcommand{\earray}{\end{array}}
\newcommand{\bea}{\begin{eqnarray}}
\newcommand{\eea}{\end{eqnarray}}

\newcommand{\rd}{\mathrm{d}}
\begin{document}
\title{Gamma ray burst delay times probe the geometry of momentum space}
\author{Laurent Freidel and Lee Smolin${}^c$  \\}

\date{\today}

\begin{abstract}

We study the application of the recently proposed framework of relative locality \cite{PRL} to the problem of energy 
dependent delays of arrival times of photons that are produced simultaneously in distant events such as gamma ray bursts.
Within this framework, possible modifications of special relativity are coded in the geometry of momentum space.   The  metric
of momentum space codes modifications in the energy momentum relation, while the 
connection on momentum space describes
possible non-linear modifications in the laws of conservation of energy and momentum.  

In this paper, we study effects of first order in the inverse Planck scale, which are coded in the torsion and non-metricity of momentum space.  We find that
time delays of order $\mbox{distance} \times \mbox{energies}/m_p$ are coded in the non-metricity of momentum space.  Current experimental
bounds on such time delays hence bound the components of this tensor of order ${1}/{m_p}$.
We also find
a new effect, whereby photons from distant sources can appear to arrive from angles slightly off the direction to the sources, which we call
{\it dual gravitational lensing.}  This is found to be coded into the torsion of momentum space.

\end{abstract}

\maketitle

\tableofcontents

\newpage
\section{Introduction}

It is a commonplace to assert that the physical consequences of the unification of quantum theory with general relativity are to be sought at tiny length scales, given by the planck length, $l_p =\sqrt{\hbar G/c^3}$. 
There is however a wealth of quantum gravitational phenomena that could manifest itself even when the planck lenght is negligible as long as the Planck mass 
\f
m_p = \sqrt{\frac{\hbar}{G_{Newton}}}
\ff
is not. We show here that these phenomena can appear at very large distance scales.

Whatever the quantum theory of gravity is, there is an experimental regime in which
$\hbar$ and $G_{Newton}$ may both be neglected, while the ratio, $m_p$ is observable.  This is then a regime of ``classical, non-gravitational",
quantum gravitational effects. 
Recently we and colleagues have  argued \cite{PRL} that in this regime one can expect a deepening of the relativity principle, called the relative locality principle which includes  a new kind of realization of translation invariance.  

A simple way to formulate theories that exhibit the relativity of locality is to posit that momentum space is curved.   This means that non-linearities may appear in both the relations between energy and momentum and in the laws of conservation of energy and momentum.  In \cite{PRL} it was shown that modifications of the energy-momentum relations are measured by the metric geometry of momentum space, while non-linearities of conservation laws are coded by the connection of the space.  We posit a correspondence principle, according to which such effects are suppressed by powers of a large mass scale, $m_{QG}$, which theory tells us to suggest may be the Planck mass, but is in fact the job of experiment to bound or measure.  Effects which appear at first order in $\frac{Energies}{m_{QG}}$ arise due to two kinds of effects: the connection geometry of momentum space may have torsion, and it may be non-metric.  Non-metricity means that the connection measured by conservation laws does not preserve the metric as measured by energy momentum relations. At second order,  $\frac{Energies^{2}}{m_{QG}^{2}}$ effects due to the momentum space curvature appear.

In these theories, the locality of physics is realized in a new way, which we call relative locality.  Observers using a coordinate system whose origin is adjacent to any interaction describe it as local, as in special relativity.  But observers using coordinates whose origin is distant from that interaction are constrained to describe it by coordinates in which the event appears to be non-local.

%

These non-localities, which can be created and eliminated by translations are, we want to stress,  not real non-localities.  They occur, not because of any subjectivity of reality, but because, as Einstein showed in his 1905 paper, the description of distant events by an observer are constructed by exchange of light signals.  In the class of theories we discuss, this coordinatization of distant events necessarily becomes energy dependent.  There remains an invariant description of physics, but this takes place in a phase space.  There is no longer an observer and energy independent notion of spacetime that may be universally projected out from the phase space.  
\begin{figure}[h!]
\begin{center}
\includegraphics[width=0.5 \textwidth]{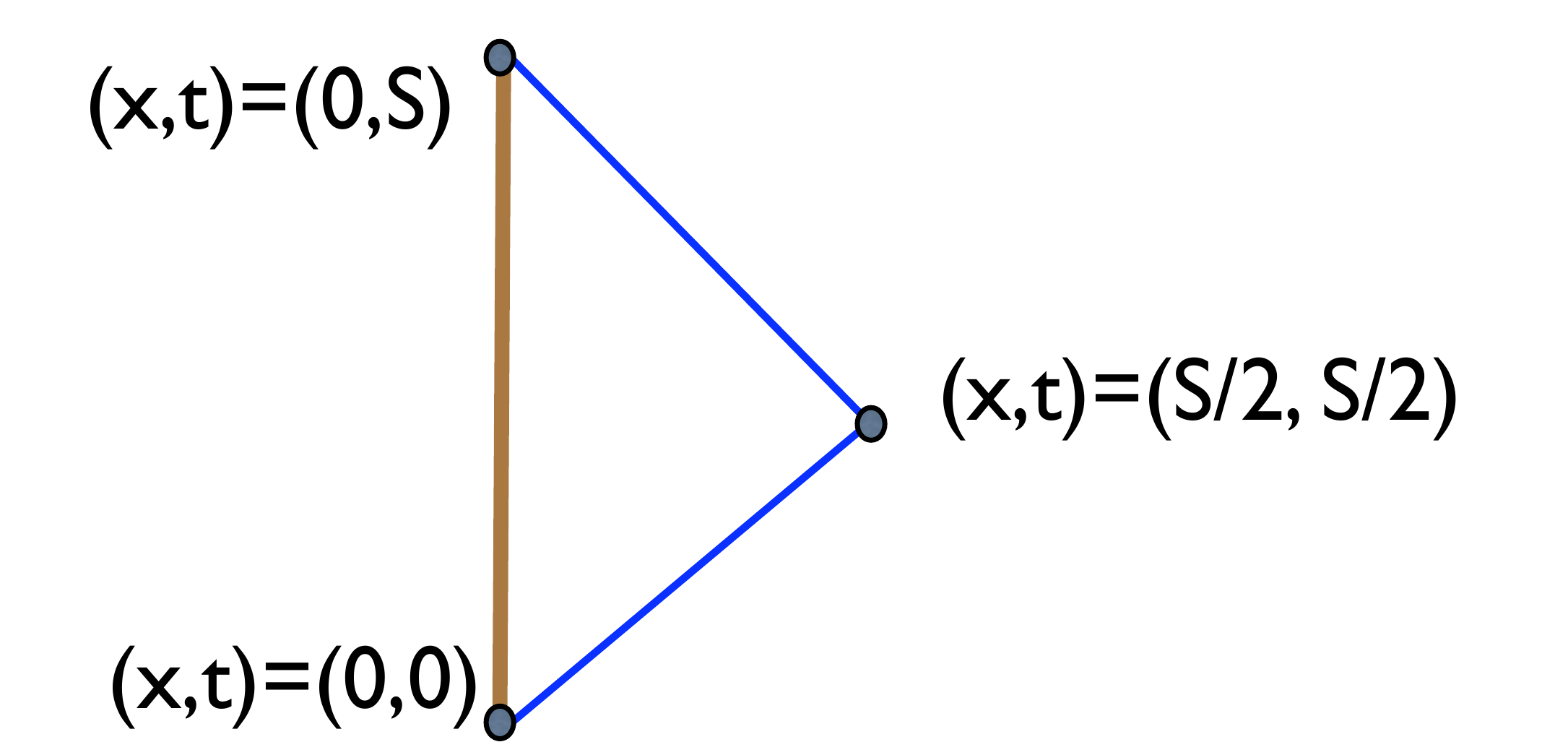}
\end{center}
\caption{The coordinates of distant events and the spacetime geometry is inferred from momentum space measurements.}
\end{figure}

Even if the apparent non-localities in a single interaction can always be removed by transforming to a coordinates system whose origin is adjacent to it, there are novel kinds of physical effects, which arise when several events, very distant from each other, are part of a single process.  In these cases, any single instance of apparent non-locality 
can be removed by transforming to coordinates whose origin is adjacent to it.  But in cases of processes involving widely separated events, any coordinate system will ascribe large values of coordinates to at least one event, which will then be subject to relative locality.   Different observers, using different coordinate systems, will attribute the non-locality to different events in the process,  but all will agree on the predictions for observable quantities. 
The challenge is to provide a clean way to understand whether such uncertainty in distant localisation really results in observable physical effects. 

In order to settle this issue we study the Einstein localisation procedure in great detail for a particular  process which is relevant 
for the description of gamma ray bursts, when photons of different energies are produced and, then detected by measurements made very distant from their production.  Think of two photons, one of very low, the other of  very high  energy, which are produced simultaneously from the point of view of an observer at the burst.  They travel together, because in these theories, as we will show, coordinates can always be chosen to preserve the universality of the speed of light. 
The question is wether they will be observed simultaneously or not given that the distortion of locality is energy dependent?

To address this question, we develop a method that is independent of particular observers and coordinates and relates physically meaningful proper times to physically meaningful energies.  This leads to two major results. 
 The first is that if there is non-metricity in the momentum space then there exists a time delay between emission and reception of photons of different energies. More precisely, if two photons are emitted at the same time we find that they are received at different times with a proper time difference given by
\be
\Delta S = \frac12 T \Delta E  N^{+++}
\ee

where $T$ is the time of flight of the photons, $\Delta E$ the difference of energies between the two photons and 
$N^{+++}$ the components of the non-metricity tensor in the photon's direction.
 This implies that current measurements already bound the tensor that measures the non-metricity of momentum space to order ${1}/{m_p}$.  
Thus, current experiments are probing the geometry of momentum space in the quantum gravity regime.  
If non metricity vanish there is no time delay effect at this order.

The second major result is  that torsion is responsible for a new kind of effect, which is a distortion of the apparent direction from which a photon from a distant source is seen to arrive.   We call this {\it dual gravitational lensing} as it arises from the curvature of momentum space rather than spacetime. More precisely, we show that if the non metricity vanishes, the apparent directions between the two photon are rotated with respect to one another by an angle proportional to the average energy time 
some component of the torsion tensor (\ref{Dtheta}).
\be
\Delta \theta = E^{av} |T_{-}^{+}|
\ee
where $|T_{-}^{+}|$ is the norm of the torsion tensor in the photon direction and $E^{av}$ is the average energy of the two photons.

There will be further effects of order $(\frac{Energies}{m_{QG}})^2$ coded by the curvature tensor of momentum space. Consideration of these is postponed to a later paper.  We note that there do exist a class of geometries which are flat, in the sense that the curvature tensor vanishes, while the torsion and non-metricity tensors remain non-vanishing.
Kappa-Poincare (see \cite{kpoinc2,kpoinc} and references therein) is an example of such a geometry.

In the next section we review from \cite{PRL} the dynamics of a system of relativistic particles interacting within the framework of relative locality.  In section III we apply this to give a very simple model of a process in which two photons are emitted and then detected, with the source and detector very distant from each other.  While oversimplified, for example, in having the same atom emit  the two photons, we believe this model suffices to compute the effects of relative locality for the gamma ray burst problem.  Section IV is required to develop some of the details of the description of interactions, which are then applied to the gamma ray burst problem  in section V.  That section contains the two main results of the time delay and dual gravitational lensing.

While this paper is largely self-contained, we recommend the previous paper \cite{PRL} as background to it.  More detail as to the mathematical framework will be presented in a forthcoming paper \cite{Lmath}.

\section{The dynamics of relative locality}

We begin the work of this paper by recalling how the dynamics of interacting particles is described within the framework of relative
locality \cite{PRL}.

\subsection{The action and basic geometry}

We associate to  any experiment an interaction diagram  representing the experimental set up.
Such a diagram represents the diagram of interaction (emission, propagation and  absorption) 
and we assign an action to such an interaction diagram.
We then describe the experiment in terms of solutions to the variational principle for this action.
\f\label{action}
S^{total} = \sum_{worldlines, I} S_{free}^{I} + \sum_{interaction, \alpha} S_{int}^{\alpha}
\ff
The free part of the action is given by
\f \label{Sfree}
S^I_{free}= \int ds \left ( x^a_I \dot{k}_a^I + { \cal N}_{\,\,\,I} {\cal C}^I (k^{I})
\right )
\ff
where $s$ is an arbitrary time parameter and ${\cal N}_{\,\,\,I}$ is the Lagrange multiplier imposing the mass shell condition
\f\label{ep1}
{\cal C}^{I}(k)\equiv D^{2}(k) -m^{2}_{I}.
\ff
The range of integration is from the initial to the final event of each worldline.

The quantity $D^{2}(k)$ is the geodesic distance from the origin of momentum space to the point labeled by the momentum $k_a$.  It is determined
by the metric on momentum space. 

The interactions are described by
\f
S_{int}^{\alpha}= - {\cal K}^{(\alpha)}_{\,\,\,a} z^a_{(\alpha)}
\label{int}
\ff
where the ${\cal K}_{\,\,\,a}^{(\alpha)}(k^{J})$ give four energy-momentum conservation laws for each interaction, which are functions of the momenta at the vertex.  
These are constructed from a non-linear combination rule, notated\footnote{The indices $a,b$ range from $0$ to $3$ and label energy and momenta} as 
\f
(p, q)
\rightarrow p^\prime_a = ( p \oplus q )_a
\ff
and also from the inversion denoted $\ominus$ which satisfy $(\ominus p) \oplus p = p \oplus (\ominus p)=0$.

This combination rule defines a parallel transport operation on the momentum space, $\cal P$, defined by
\f
p_a \oplus dq_a = p_a + U(p)_a^b dq_b = p_a + dq_b  + \Gamma_a^{bc}p_b dq_c + ...
\ff
where $U_b^a$ is a parallel transport operator whose infinitesimal expression is a connection evaluated at the origin $\Gamma_a^{bc}$. Notice that the indices are all reversed
in positions, because we are working on a space of momenta;  here the indices on $p_a$ are coordinate
and tangent space indices, but we keep the convention of coding momenta with lower indices. 

More complicated interactions are built up of combinations of these, for example the conservation law of a three body interaction might be written down 
as,
\f
{\cal K}_{\,\,a}^{(3)}(q,p,k)_a = ((q \oplus p) \oplus (\ominus k))_a =0
\ff 
where incoming momenta $p$ appear as such in the non linear  addition while the outgoing momenta $k$
appears with the inversion $\ominus k$.
Neither commutativity or associativity is assumed, these are, as shown in \cite{Lmath}, measured by the torsion and curvature of the connection.

Before going further we have to point out some subtleties of the geometry, which are a consequence of the curvature of momentum space $\cal P$. 

\subsection{The structure of the phase space}
\begin{figure}[h!]
\begin{center}
\includegraphics[width=0.5 \textwidth]{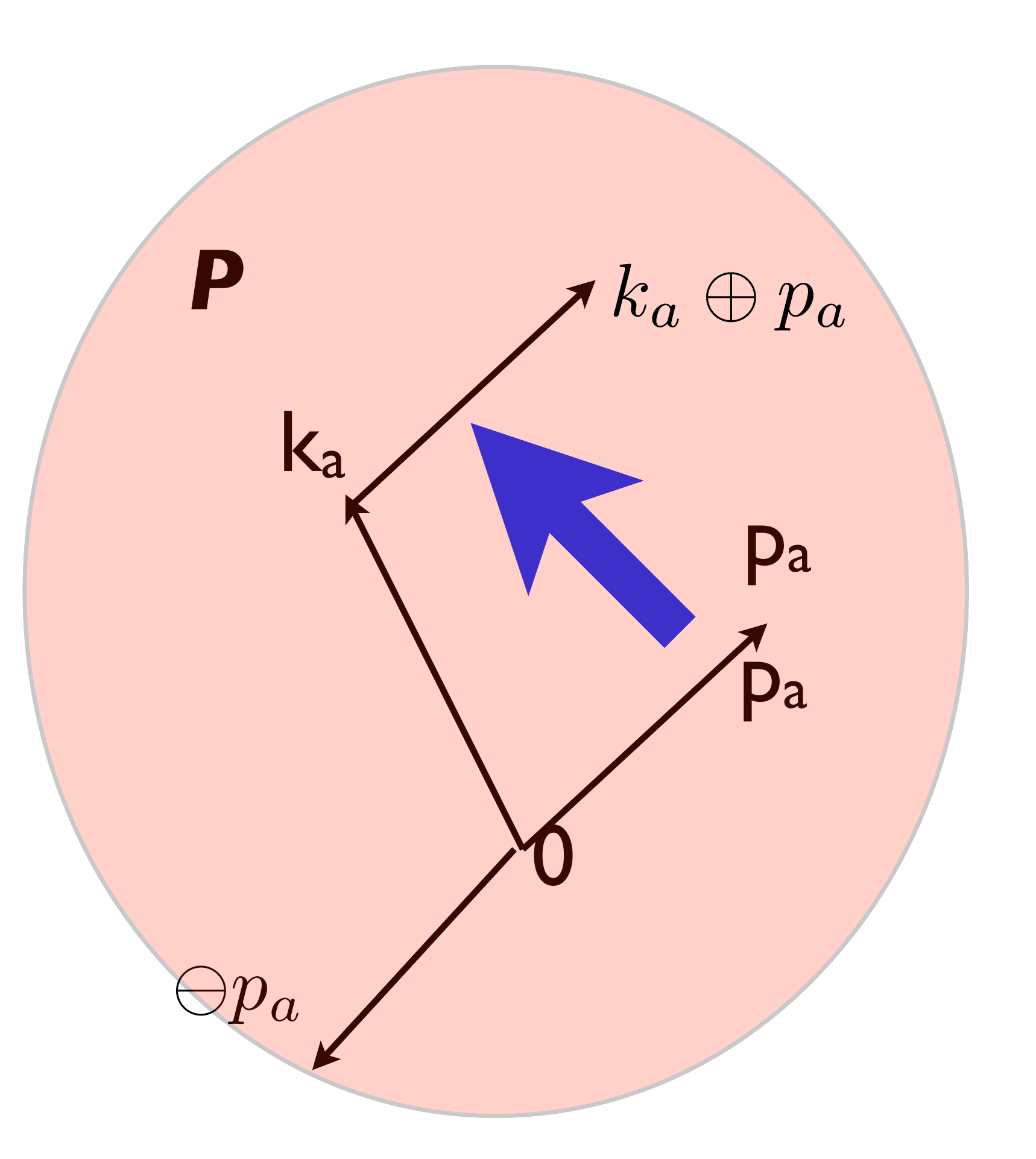}
\end{center}
\label{momentum space}
\caption{The momentum space showing the parallel transport defined by the non-linear combination of momenta.}
\end{figure}

First, notice that the metric of spacetime appears nowhere in the action. Only the geometry of momentum space comes in.  

In the dynamics defined by the action (\ref{action}), the spacetime coordinates are auxiliary quantities that are, in a sense, emergent
from the more fundamental momentum space data.  The spacetime geometry is yet to be constructed from the solutions of the equations of motion.

To understand the structure of the phase space it is important to emphasize that there are two kinds of spacetime coordinates that appear in the 
action.  There are the $x^a$'s in the free term and the $z^a$'s in the interaction terms.  There is one $x^a$ for each worldline and one
$z^a$ for each interaction vertex.  These two kinds of coordinates have very different geometric meanings and roles and it will be important
to distinguish them.  

We see that in (\ref{Sfree}), the coordinates of the free particle trajectories $x^a$ arise as conjugate variables to the coordinates of the momenta.  
That is, they satisfy,
\f\label{canonical}
\{x^a , k_b \} = \delta^a_b
\ff
The pairs $(k_a, x^a)$ then coordinatize the phase space $\Gamma$.  What is the structure of this phase space?

Because the momentum space, $\cal P$ is curved, the phase space is the cotangent space of $\cal P$.  
\f
\Gamma = T^{*} ({\cal P})
\ff
Geometrically, the $x^a_I$'s  live in the cotangent space of the momentum space $\cal P$, at the point of the corresponding conjugate momenta, $k_a^I$.   
That is, the $x^a$ defined by (\ref{Sfree}) and (\ref{canonical}) is implicitly dependent on the $k_a$, as it coordinatizes the cotangent plane to $\cal P$ at the point $k_a$.

The fact that momentum space is curved means
that the coordinates of particles with different momenta live in different cotangent spaces.  
So the worldline $x^a(s)$ that results from solving the equations of motion lives in the cotangent plane over the momentum $k_a$ that is carried by the worldline.  
\begin{figure}[h!]
\begin{center}
\includegraphics[width=0.5 \textwidth]{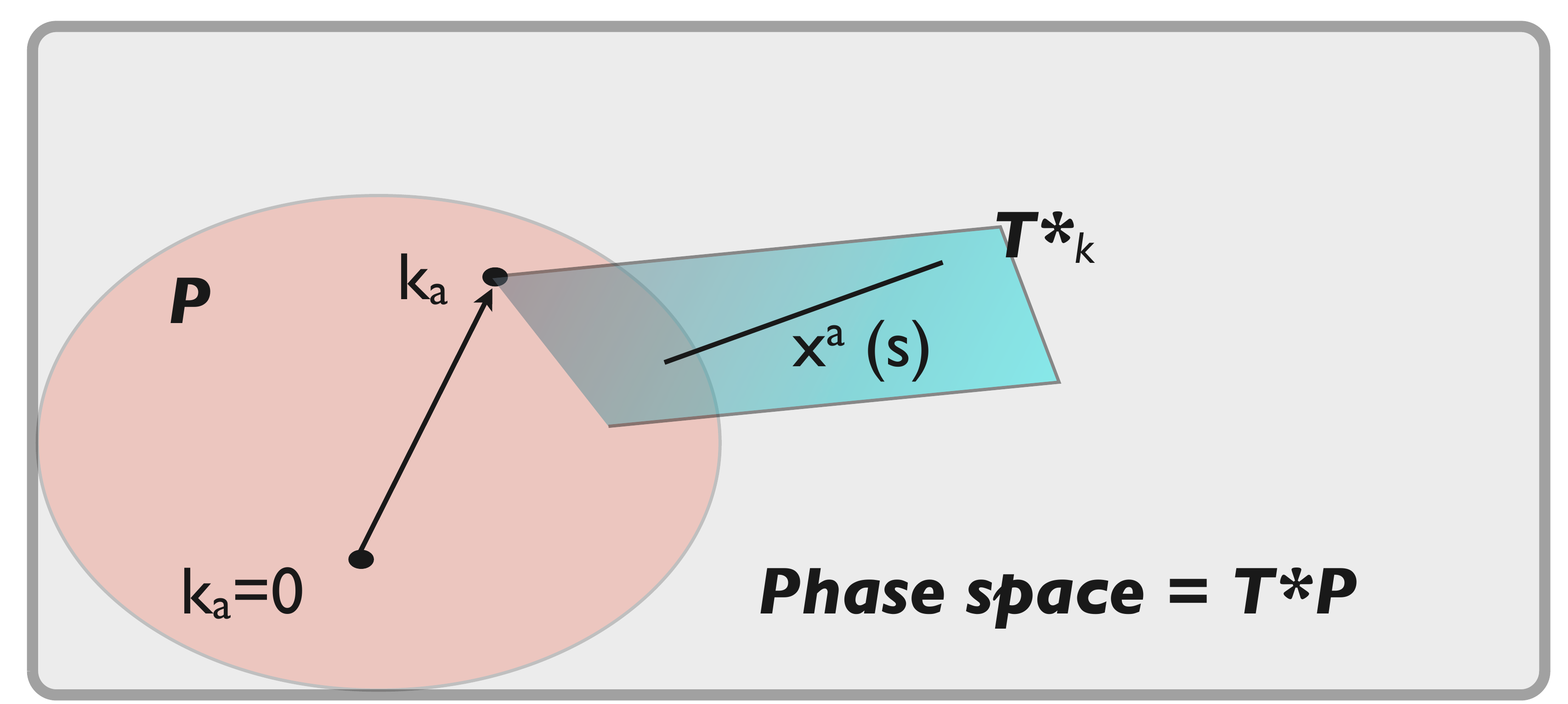}
\end{center}
\label{cot1}
\caption{The coordinates $x^a$ conjugate to $k_a$ live in the cotangent space of momentum space at the point $k_a$.  Worldlines with different
momenta live in different spaces.  The whole phase space is the cotangent bundle of momentum space $\Gamma= T^{*} ({\cal P}).$}
\end{figure}

Let us consider then an interaction, where three worldlines are to interact, each carrying a different momentum, $k_a^I$, with $I=1,2,3$. 
These live in three different cotangent planes, $T^{*} (k_a^I).$ To describe how they interact we need to parallel transport them to
the same cotangent space, where we can add them.  The absolutely key point that underlies relative locality is that this is necessary.
There is no invariant meaning to adding vectors in different spaces, so the best we can do is to parallel transport them to the same space
and add the resulting vectors there.  

The interaction terms will  tell us how to accomplish this parallel transport.  But before studying that we need to spend a minute on the 
role of diffeomorphism invariance in the dynamics.

\subsection{Invariance under diffeomorphisms of momentum space}

As stressed in \cite{Lmath}
we require that physical observables are invariant under diffeomorphisms of the momentum space, $\cal P$, which may be denoted
\f
k_a \rightarrow k_a^{\prime} = F_a (k)
\ff
under these diffeos, elements of the tangent space at $k_a$ transform as
\f
dk_a \rightarrow dk_a^{\prime} = \frac{dF_a}{dk_b}dk_b|_{k}
\ff
The free action is invariant under these diffeomorphisms so long as the $x^a_I$'s transform like covectors
\f
x^{a \prime}|_{k} = \left [ \frac{dF}{dk} (k)  \right ]^{-1 \  a}_b x^b
\ff
Note that this means that diffeomorphisms on momentum space $\cal P$ induce transformations on the spacetime coordinates that generally mix them up with momenta.  This is what we mean when we say that $x^{a}$ implicitly dependent on the momenta and when we emphasize that there is no invariant meaning of spacetime, apart as momentum dependent sections of phase space.  

The interactions part   also carry an invariance under diffeomorphisms of $\cal P$. But since the ${\cal K}_a =0$ by the equations of motion, the
interaction coordinates $z^a$ transform in the cotangent space of the origin of momentum space, $k_a=0$.  The equation of motions are invariant under transformations
induced by diffeomorphisms of $\cal P$,  if
\f
z^{a \prime}| = \left [ \frac{dF}{dk} (k=0)  \right ]^{-1 \  a}_b z^b
\ff
\begin{figure}[h!]
\begin{center}
\includegraphics[width=0.5 \textwidth]{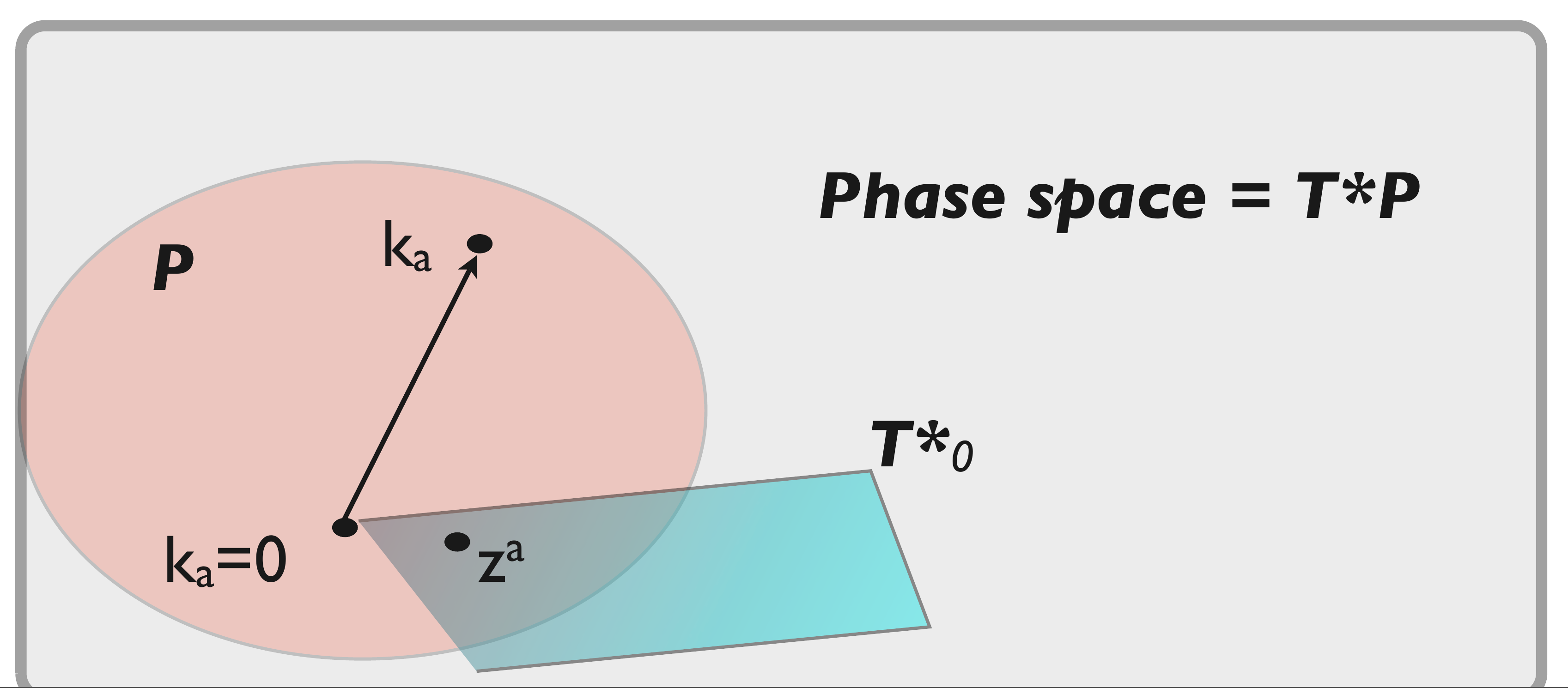}
\end{center}
\label{cot2}
\caption{The coordinates $z^a$ live in the cotangent space of momentum space at the origin of momentum space.}
\end{figure}

\subsection{The equations of motion}

Let us write down the equations of motion that come from varying the action. 
First, there are the equations of motion that come from varying the free parts of the action which define the worldlines,
\begin{eqnarray}
\dot{k}_a^I &=& 0
\nonumber \\
\dot{x}^a_I &=& {\cal N}_{\, \,\,I} \frac{\delta {\cal C}^I}{\delta k_a^I}
\nonumber \\
{\cal C}^I (k) &=&0   
\label{EOM1}
\end{eqnarray}

Then there are two equations of motion that come from the interaction terms.  Varying the $z^a_{(\alpha)}$'s we get the conservation laws for each vertex
\f\label{cons}
{\cal K}_{\,\,a}^{(\alpha)} =0
\ff
The most important equation is that from varying the momenta $k_a (0)$ at the endpoints of the worldlines, as these come into both the
free and interaction terms of the interaction.  These are what relates the $x^a_I$'s to the $z^a$'s.  To be precise, they relate the 
$x^a$ at each endpoint of a worldline at a vertex with the $z^a$ at that vertex.  These crucial relations have the form
 \f
\boxed{ x^a_I (0) = z^b ({ \cal W}_{x_{I}})_b^a  
}
\label{parallel}
\ff
where the linear transformations ${{\cal W}}_b^a$ are given by the matrices
\f\label{defparalell}
{ ({\cal W}_{x_{I}})}_b^a = \pm  \frac{\delta {\cal K}_{\,\,b}}{\delta k_a^I }.
\ff
Here the $+$ sign appears for incoming momenta, while $-$ sign appears for outgoing momenta.  These signs 
distinguish an initial from a final point when integrating by part, moreover the convention chosen for incoming and outgoing momenta in the interaction are such that 
for small momenta ${ ({\cal W}_{x_{I}})}_b^a  \approx \delta_b^a$.  

The equations (\ref{parallel}) are the essence of relative locality, and we will call them the
{\it relative locality relations}.   They express the fact that {\it the interactions are as local as possible,
given that the wordlines and interaction coordinates live in different spaces.}  It would be non-invariant
non-sense to just set the $x^a(0)$'s equal to the corresponding $z^a$.  What   (\ref{parallel}) says is that
{\it parallel transported to the same spaces, the $x^a(0)$'s  become equal to the corresponding $z^a$.}

To define a parallel transport on a curved space you need two ingredients.  One is a connection, which is given
by the geometry of momentum space.  The other is a path along which to parallel transport.  That is, as we will
see in detail below, given by the conservation law at each vertex.  

\begin{figure}[h!]
\begin{center}
\includegraphics[width=0.5 \textwidth]{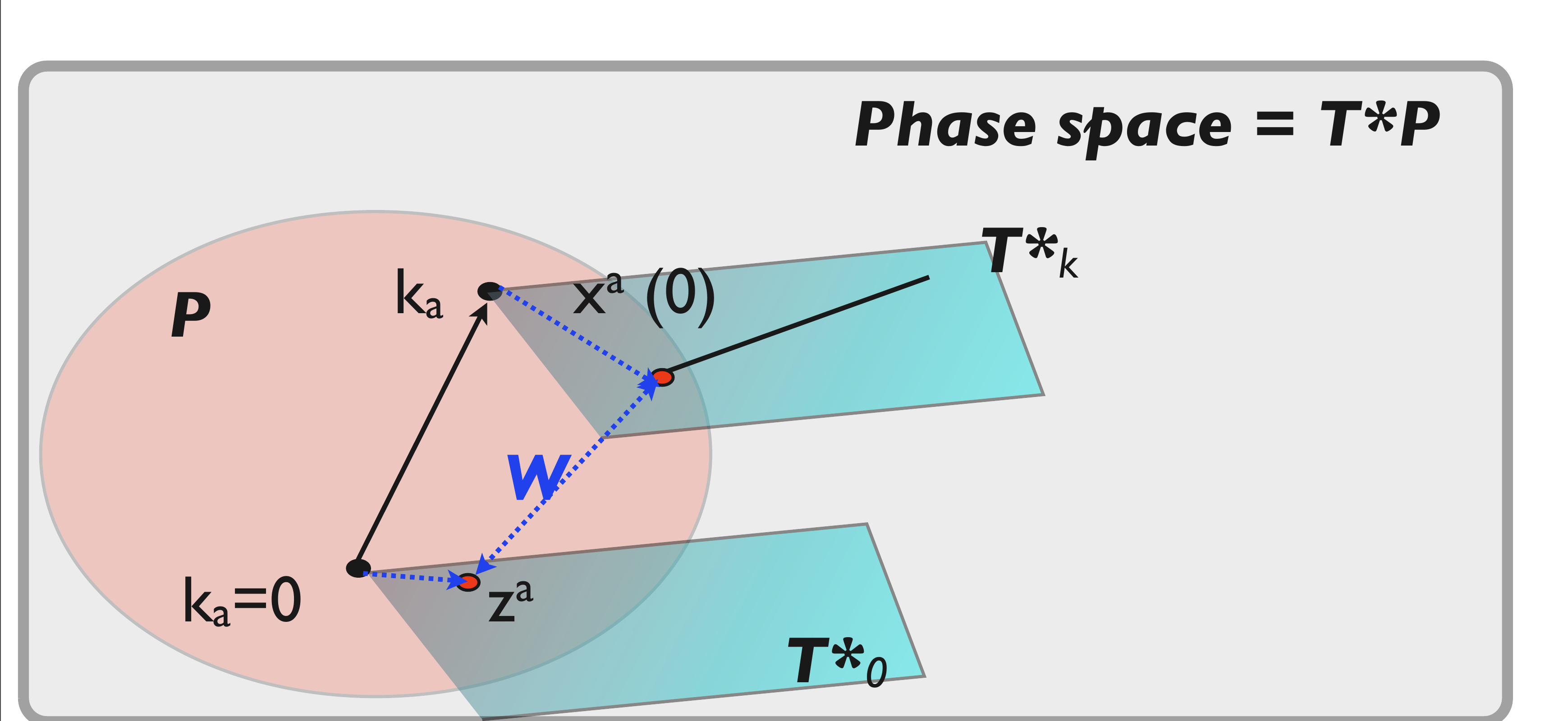}
\end{center}
\label{rellocrelation1}
\caption{Illustration of the relative locality relations (\ref{parallel}).}
\end{figure}

Note that this has several consequences that can be read off of the relative locality relations (\ref{parallel}).

\begin{enumerate}

\item{}When energy-momentum conservation is linear then ${\cal W}_b^a=\delta_b^a$ and all the $x^a(0)=z^a$,
so that we have the usual picture in which interactions are local in special relativity.  Thus, the usual notion of locality is
a consequence of the assumption that energy momentum conservation is linear.  

But what happens when energy momentum conservation is non-linear?  Then ${\cal W}_b^a \neq \delta_b^a$ and the differences
between the $x^a(0)$ and the $z^a$'s at the same vertex are momentum dependent.

\item{}In the coordinates of an observer for whom $z^a=0$ it is also the case that the $x^a (0)=0$.  This is to say
that {\it observers for whom the interaction takes place at the origin of their coordinates see the interaction to be
local. }  This is the sense in which locality is respected.  For every interaction there are observers who see it to be local.

\item{}In the coordinates of an observer distant from the interaction, so $z^a \neq 0$, it follows that in general
$x^a(0) \neq z^a$.  So in the coordinates used by distant observers it appears as if interactions are non-local.  This does not mean
the theory has physical non-localities, because the non-locality of any single interaction can be made to be local by transforming to
an observer adjacent to the interaction who will see it to be local. 

\item{}However, in a process involving interactions separated by long distances it is not possible for any single observer to remove all the 
non-locality effects by choice of coordinates.  In this case there can be real physical consequences of relative locality, as we will see.  
However where the non-locality occurs will depend on the observer, it is always only at events distant from the observer. 

\end{enumerate}

\begin{figure}[h!]
\begin{center}
\includegraphics[width=0.5 \textwidth]{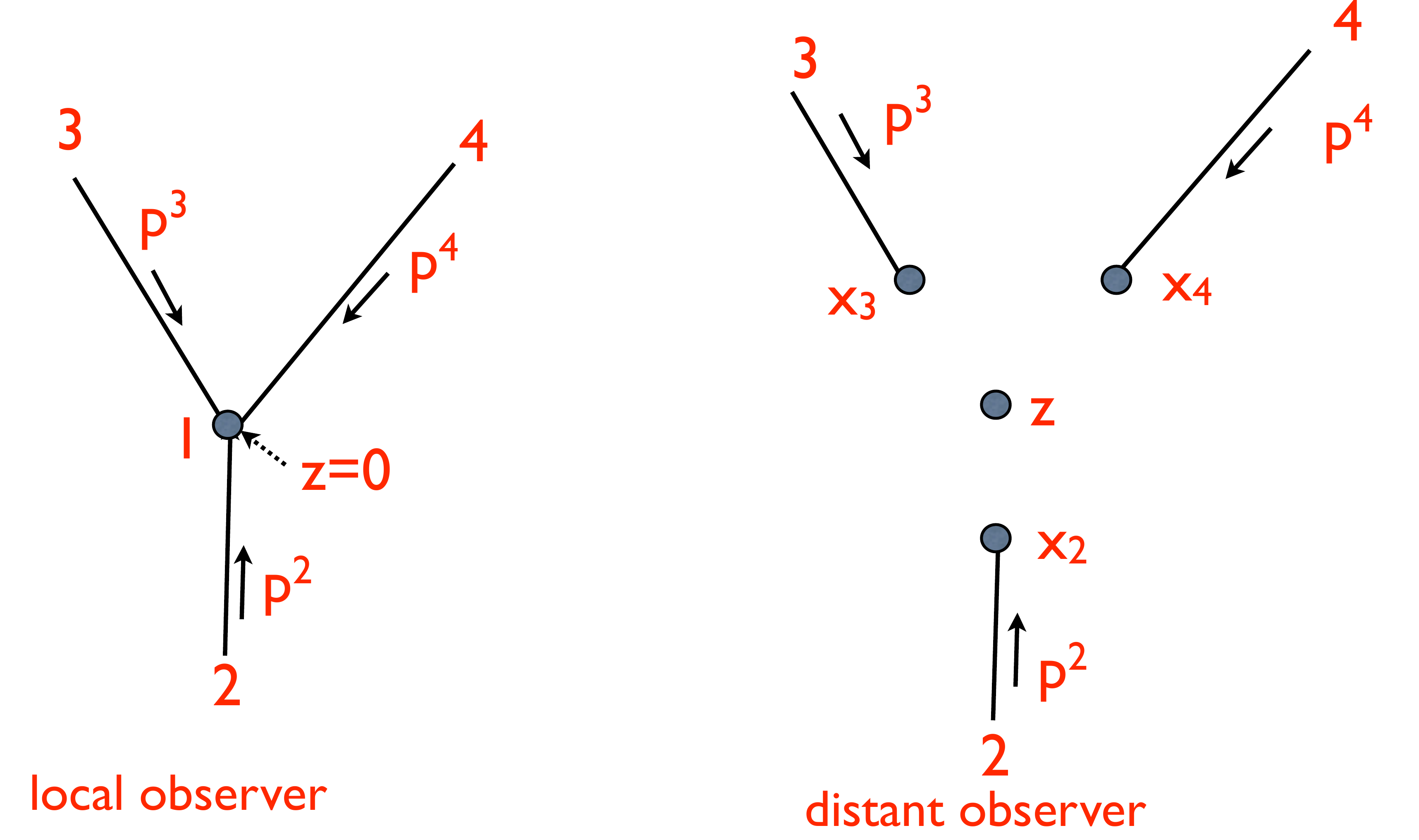}
\end{center}
\label{relloc1}
\caption{When projected from phase space into spacetime coordinates, a  local interaction appears local to observers local to it, but appears spread out using the coordinates of distant observers.}
\end{figure}

\subsection{How interactions define parallel transport}

We can now explain how the interaction terms define the parallel transport between the worldlines in different cotangent planes of the phase space. 

Let us look at the two equations of motion that follow from varying the degrees of freedom in the 
interaction terms.  The first (\ref{cons}), tells us that the conservation law ${\cal K}_{\,\,a}$ vanishes.  This means that the corresponding $z^a$ lives in the cotangent
plane over the origin of momentum space,
\f
z^a \in T^{*} (0)
\ff
The $z^a$'s come in as  Lagrange multipliers 
whose variation yields the conservation laws.   They become the "interaction coordinates", which are where the interation takes place in spacetime according to observers local to it. 

The second equation (\ref{parallel}) tells us the relation between the interaction coordinate, $z^a$ of a vertex and the $x^a_I (0)$ at the endpoint of a worldline
that begins at that vertex. These live, respectively in the cotangent plane at $k_a^{I}$, the momentum on that worldline, and the cotangent plane at the origin.
The ${\cal W}_b^a$ in (\ref{parallel}) are parallel transport operators that relate vectors in these two linear spaces.  Notice the crucial point that the choice
of these parallel transport operators is defined by the form of each interaction vertex.  It is not just a property of the geometry of momentum space, it has information about the dynamics.

\subsection{Riemann normal coordinates}

We will below find it convenient to use Riemann normal coordinate \cite{Lmath}.  But there is a subtelty that will concern us which is that in general there
are two kinds of  normal coordinates.  The first are {\it connection  normal coordinates.}  These allow us to set the symmetric part of the connection coefficients
to zero at a point. This means that the connection is just given by the torsion in these coordinates.   We will take the preferred point to be the origin of momentum space, which is an invariant.  When these have been imposed we have
\f
\Gamma_a^{bc}(0) = \frac12 T_a^{bc} 
\ff
where $ T_{a}^{bc}= 2\Gamma_{a}^{[bc]}$ are the torsion coefficients.
There are also {\it metric or Riemann normal coordinates.}  These set the Christoffel coefficients to zero.  
\f
\{{}_a^{bc} \} =0
\ff
These imply that the distance function from the origin to any point $k_a \in {\cal P}$ is quadratic.
\f
D^{2}(k) = \eta^{ab} k_a k_b
\ff
We emphasize the distinction between the two definitions of Riemann normal coordinates, because when the non-metricity tensor is non-vanishing they are not the same.

\section{A model of the gamma ray time delay problem}

\begin{figure}[h!] 
\begin{center}
\includegraphics[width=0.5 \textwidth]{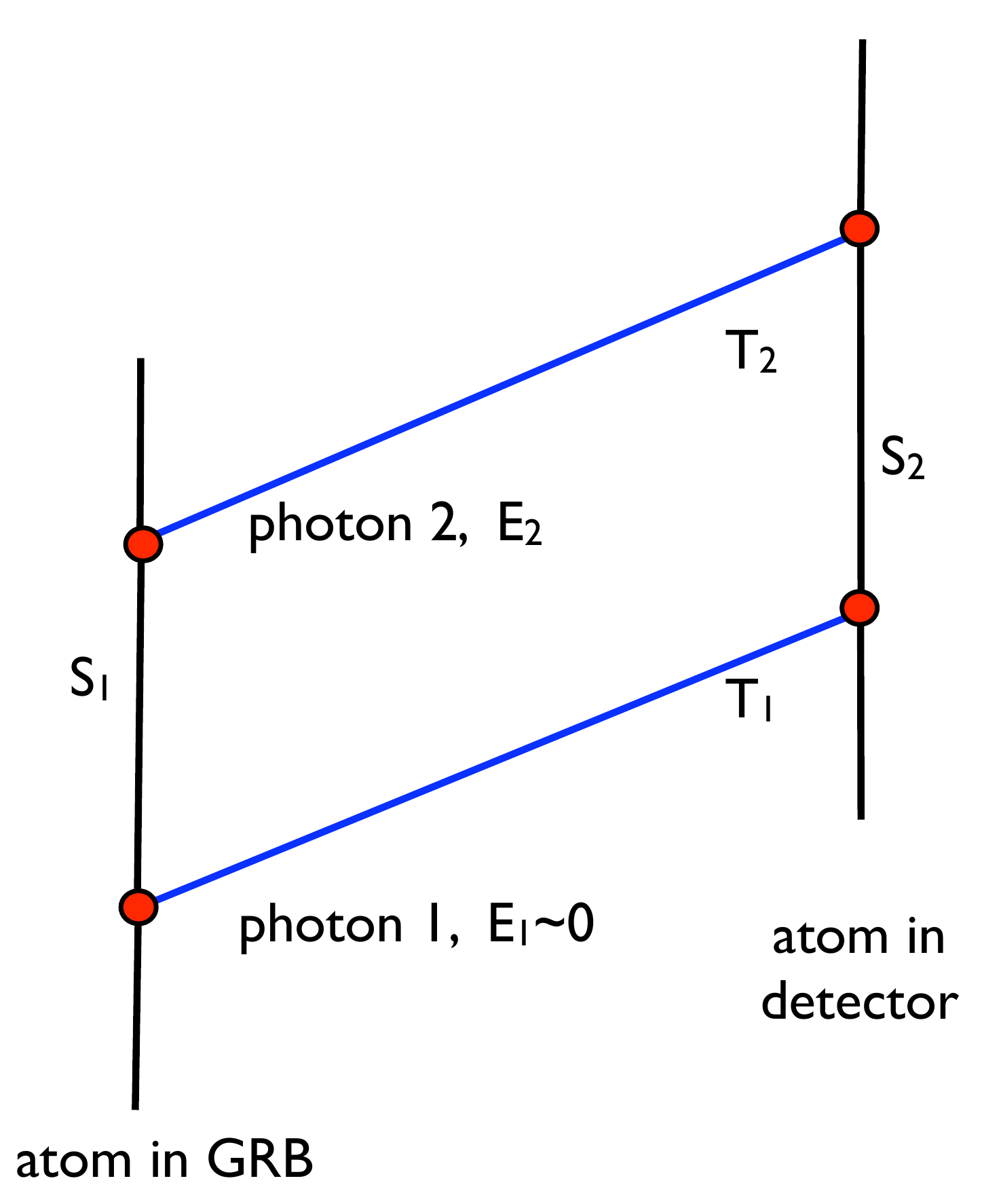}

\end{center}
\caption{\label{setup} An invariant description of the time of flight measurement}
\end{figure}

Let us consider the experiment displayed in Figure (\ref{setup}).   This is a model of an observation where two photons of different energies are emitted by a gammay-ray burst and, after traveling for a very long time, $T$, arrive at a detector.  We will include the case where the photons are emitted simultaneously, 
in an invariant sense.   This will be clearer if we begin with the case in which there are delays between the emissions and the detection of the two photons, and then take the limit where the invariant proper time between the emission events is taken to zero. 

\subsection{The basic experimental set up }

\begin{figure}[h!]
\begin{center}
\includegraphics[width=0.5 \textwidth]{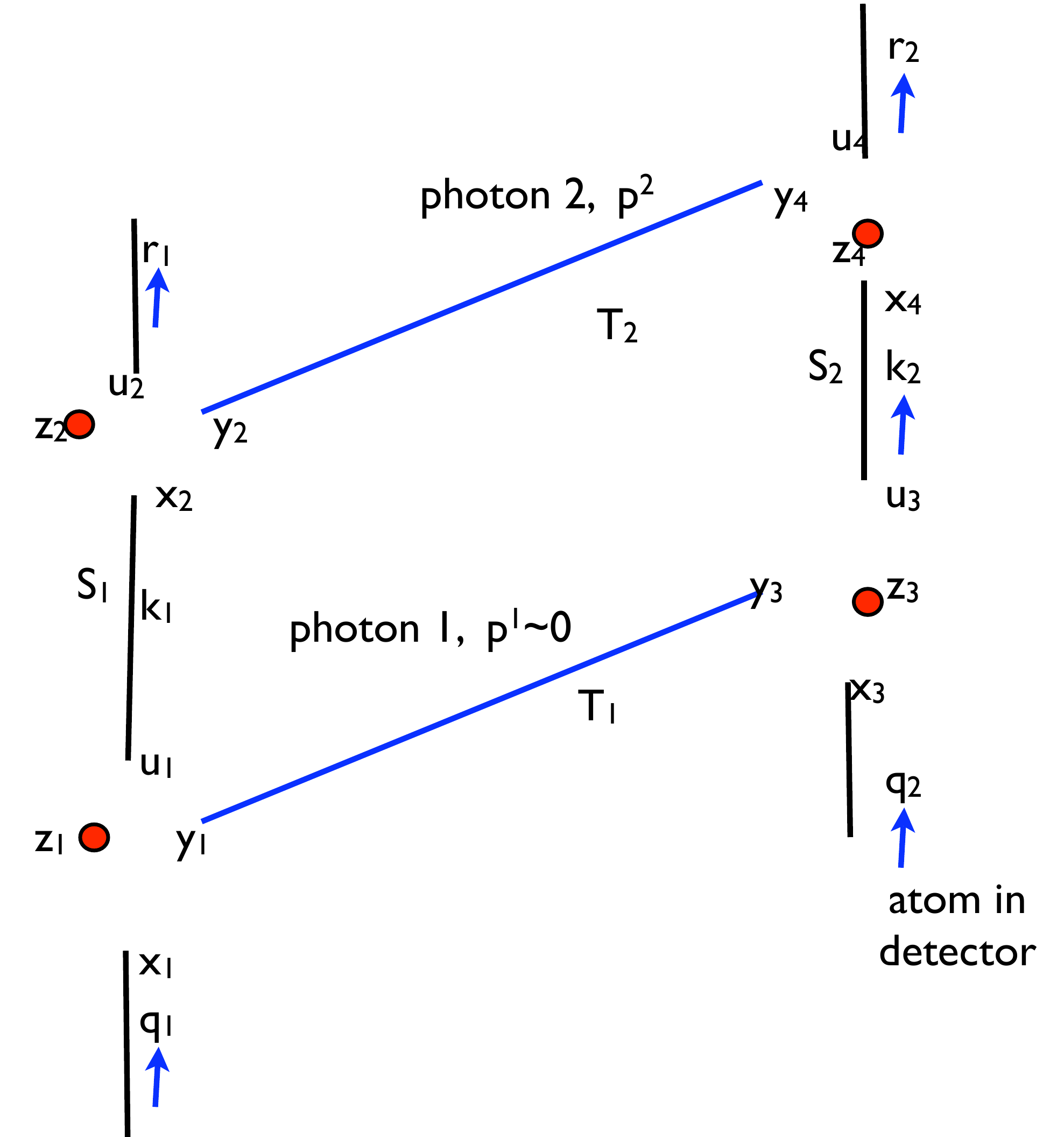}
\end{center}
\caption{Labels of positions and momenta for our model of GRB experiment. \label{fullsetup}}
\end{figure}

As shown in Figure  (\ref{fullsetup}) an emitter with initial four momentum $q_a^1$ emits a photon at an event $E_1$ with a small
momentum $p_a^1$, leaving the detector with momentum $k_a^1$.  This occurs at an interaction coordinate $z_1^a$,
the momentum is transfered from the emitter at position $x_1^a$, and the new world-line of the detector begins at 
$u_1^a$.  The photon is created at $y_1^a$.   After a proper time $S_1$ in the frame of the emitter, it emits a second,
and more energetic photon, with momentum $p_a^2$ at an event $E_2$.  The interaction coordinate of this second
emission is $z_2^a$,
the momentum is transfered from the emitter at position $x_2^a$, and the new world-line of the detector begins at 
$u_2^a$.  The photon is created at $y_2^a$. The detector is left with momentum $r_a^1$.

The first photon travels for a time $T_1$ in the rest frame of the emitter, at which time it is absorbed by a detector, which is at 
rest with respect to the emitter and has initially momentum-energy, $q_a^2$ and, after the detection, momentum $k_a^2$.  
This absorption of the first photon by the detector
is event $E_3$.  The interaction coordinate of this third
emission is $z_3^a$, the photon is absorbed at event $y^a_3$, the detector jumps from position $x_3^a$ to 
$u_3^a$ when it absorbs the photon.  

Then, after a proper time $S_2$ in the state $k_a^2$, the detector
absorbs the second photon at a fourth event, $E_4$.  This leaves the detector with momentum $r_a^2$.  The fourth
interaction coordinate associated with the detection of the second photon is $z_4^a$, the photon dissapears at
event $y_4^a$, the detector jumps from coordinate $x_4^a$ to $u_4^a$ when its momentum changes from $k_a^2$ to
$r_a^2$.  

The time the second photon traveled was $T_2$.

\subsection{Definition of the proper times}

We will need expressions for the proper times of propagation of the atoms and photons.  These are the invariant quantites that characterize
the time delays between emission and detection of the photons.

 To evaluate these we pick Riemann normal coordinates so the speed of light is universal.
 In these coordinates the velocity is simply related to the momenta by $\dot{x}=p$.
 The end points of the emitter/detector  trajectories are then related to the momenta by
 \f
x_2^a - u_1^a = \hat{k}_1^a S_1, \ \ \ \ \ \ \ \ \ x_4^a - u_3^a = \hat{k}_2^a S_2,
\label{atompt}
\ff
where $\hat{k}_{i} =k_{i}/m_{i}$ and $S_{i}$ are by definition the proper times.

The end points of the photon trajectories are then related to the momenta by
\f
y_3^a - y_1^a =   \hat{p}^a_1 T_1 ; \ \ \ \ y_4^a - y_2^a = \hat{p}^a_2 T_2
\label{gammapt}
\ff
 where $\hat{p}$ are the normalised null momenta with respect to the emitter frame and satisfy
 \be
 \hat{p}_{i}\cdot \hat{k}_{1}=1
 \ee

\section{The geometry of interactions and the mechanics of relative locality}

Before doing the computation of the time delay effect we have to understand the detailed geometry and physics at the interaction vertices.
 
  We will start with some basic definitions, then begin our study of interactions with two valent nodes.  This will help us to fix notation and understand the concepts better.
This will equip us to understand the three valent interactions that come into the emissions and absorptions of photons.

\subsection{The elementary parallel transport operations}

The non-linear combination rule for momenta is of the form $(p \oplus q)_a$.  We note that this can be 
expressed in terms of left and right translation operators on the momentum space.  These are defined as 
\f
p \oplus q = { L}_p  (q) = { R}_{ q} (p)
\ff
Note that we sometimes will suppress the indices, when the context is clear. 
We need to understand how to take the derivatives of these operations.  Let us start with the left translation operator. ${ L}_p  (q): {\cal P} \rightarrow {\cal P}$.  Its derivative defines a   parallel transport operators $(U_{p}^{q})_a^b : T_q \rightarrow T_p$, where $T_{p}$ denotes the tangent space of momentum space at the point p. We call this the left handed parallel transport operator.  It satisfies
\f\label{Ltransport}
(U^{q}_{p \oplus q} )_a^b \equiv \frac{\partial (p\oplus q)_a}{\partial q_b}=  (\rd_{q}L_{p})_{a}^{b}
\ff
where $\rd_{p} F$ denotes the differential of $F$ at the point $p$.

The  right handed parallel transport operator,  $(V_{p}^{q})_a^b : T_q \rightarrow T_p$  is similarly defined as 
\f\label{Rtransport}
(V^{p}_{p \oplus q} )_a^b \equiv \frac{\partial (p\oplus q)_a}{\partial p_b}=  (\rd_{p}R_{q})_{a}^{b}
\ff
There is also an inverse operation defined on momentum space
\f
p \rightarrow \ominus p
\ff
This turns incoming into outgoing momenta.
It satisfies
\f
 ( \ominus p )\oplus p =0
\ff
and, more generally,
\f
(\ominus p )\oplus (p\oplus k)= k.  
\ff

We will need also to  introduce the differential of the inverse operation
\be
(I^{p})^{b}_{a}\equiv   \frac{\partial (\ominus p)_a}{\partial p_b} =\rd_{p} \ominus
 \ee

These three parallel transport operations, $(U^{q}_{p \oplus q} )_a^b$, $(V^{p}_{p \oplus q} )_a^b$ and $(I^{p})^{b}_{a}$ are geometric.  They are the basic building blocks that the parallel transport operators defined
in (\ref{parallel}) by the dynamics are made up of.  

These three basic operators are not independent, they are related by a basic identity
\f
\label{binode}
-U_{0}^{\ominus p}I^{p} = V_{0}^{p}
\ff 
Before showing this, we can check it to leading order in the non linearity.
  To do this, we can use the definition of the connection coefficients, 
\f
\Gamma_a^{bc} = -\frac{\partial}{\partial p_c} \frac{\partial}{\partial q_b} (p_a \oplus q_a)|_{p_a= q_a =0}
\ff
to write for small momenta
\f
(p \oplus q)_a = p_a + q_a - \Gamma^{bc}_a p_b q_c + \cdots 
\ff
to this  order we also have
\f
(\ominus p)_{a} = -p_{a} - \Gamma^{bc}_{a}p_{b}p_{c} +\cdots
\ff
Note that it is always possible to define normal coordinates for the addition such that 
\f
\ominus p = -p
\ff
to all order in $p$.

Doing the computation to first order in the non linearity we find
\f
(V_{0}^{p})_{a}^{b} = \delta_a^b  + \Gamma_a^{bc} p_c + ...
\ff
\bea\nonumber
(U_{0}^{ \ominus p} I^{p})_a^b &=& (\delta_a^d  - \Gamma_a^{cd} p_c) \left( -\delta_{d}^{b} - (\Gamma^{bc}_{d} + \Gamma^{cb}_{d})p_{c} \right)=-V^{p}_{0}
\eea
so the desire identity (\ref{binode}) is satisfied to leading order. 

We can show that this is true to all orders:
by taking the derivative with respect to $p$ of the left inverse property and using the chain rule of derivation 
we find
\be
0=\partial_{p} (\ominus p \oplus (p\oplus q))_{\mu}= U_{q}^{p\oplus q}V_{p\oplus q}^{p} + V^{\ominus p}_{q}I^{p}
\ee
specialising to the case $q=0$ we obtain (\ref{binode}).

\subsection{The structure of two valent nodes}

\begin{figure}[h!]
\begin{center}
\includegraphics[width=0.5 \textwidth]{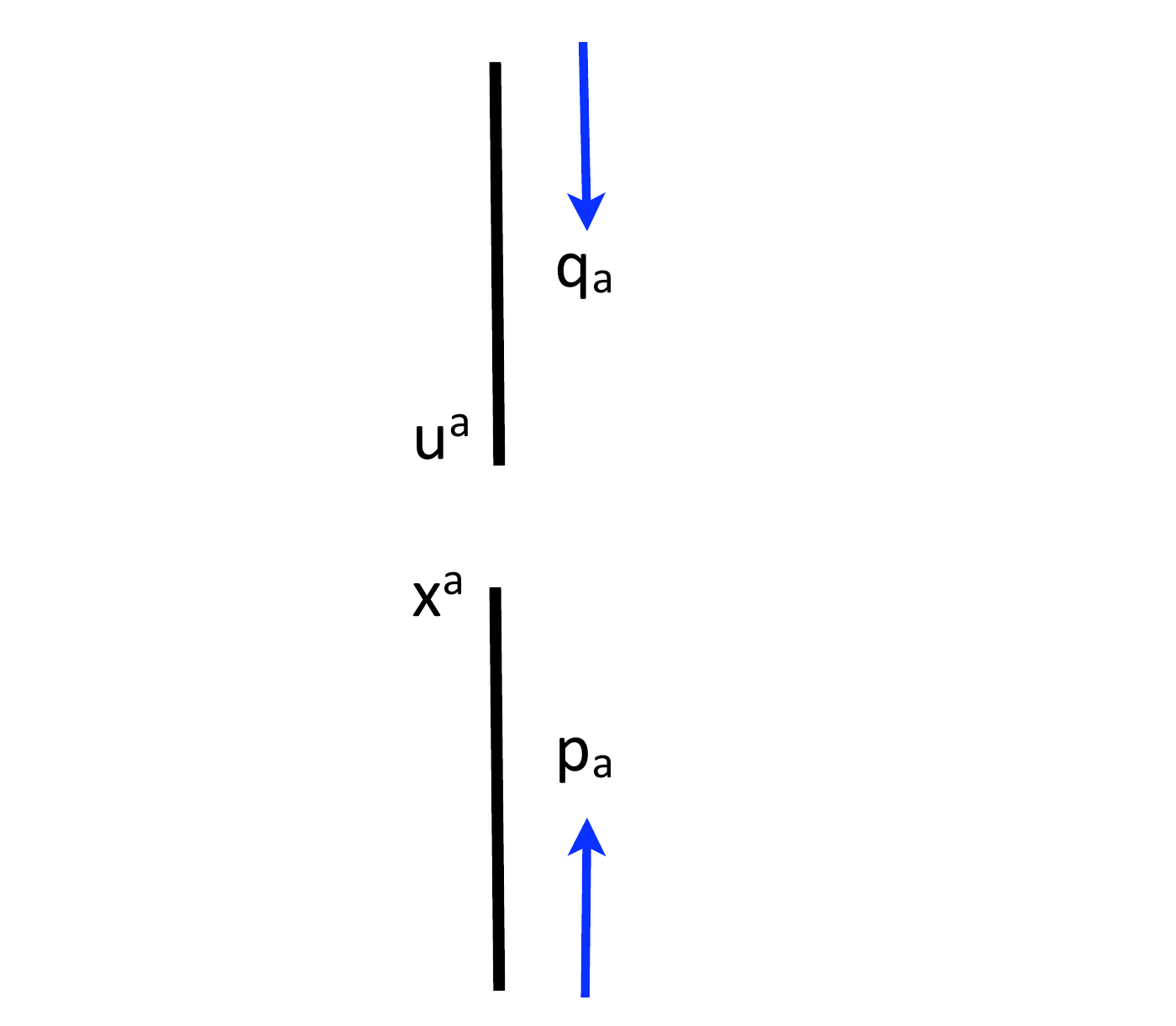}
\end{center}

\caption{\label{twovalent1} A two valent node}
\end{figure}

As a warm up we will ask a simple question: does relative locality afflicts even two valent nodes?   This will give us some experience with the kind of 
calculations we need to do to understand more complicated interactions.  What we particularly need to understand is how the parallel transport defined by
the conservation laws at vertices is composed of the elementary geometric parallel transport operators.  This is the same as understanding how the conservation laws define paths in momentum space along which the connection is integrated to give the parallel transport matrices, ${\cal W}_b^a$.  

We consider a two valent node, with momenta $p_a$ incoming from the past to an endpoint, $x^a$ and momenta $q^a$ outgoing to the future with starting point
$u^a$. This is shown in Figure (\ref{twovalent1}).  The conservation law is
\f
{\cal K}_{\;\,a}^{(2)} = (p \ominus q)_a =0
\label{p+q}
\ff
where we have denoted $p\ominus p$ for $p\oplus (\ominus q)$.
We note that
\f
{\cal K}^{(2)} = p \ominus q = { L}_p ( \ominus q) = { R}_{\ominus q} (p)
\ff

Suppose that this is implemented by an action which includes the interaction term
\f
S^{int}= -z^a {\cal K}^{(2)}_{\,\,\,a}
\ff
The equations of motion will include, in addition to (\ref{p+q}),  the relative locality conditions (from \ref{parallel}), 
\f
x^a = z^b  \frac{\partial {\cal K}_{\;\,b} ^{(2)}}{\partial p_a}  =   z^b  (V_{p \ominus q}^p)_b^a.
\ff
\f
u^a = -z^b   \frac{\partial {\cal K}_{\;\,b}^{(2)}}{\partial q_a}  =  -z^c (U_{p \ominus q}^{\ominus q})_c^{b} (I^{q})_{b}^{a} 
\ff

We see here that the parallel transport operators defined by the vertex are made up of the elementary geometric parallel transport operations.

In the following we use a short-hand for index contraction such that 
$(z U V)^{a}\equiv z^{c} U_{c}^{b}V_{b}^{a}$.
The solution to (\ref{p+q}) is
\f
q= p
\ff
So we have 
\f
x = zV_{0}^{p} ,\qquad u = -z U_{0}^{\ominus p}I^{p}
\ff

We see that the basic relation (\ref{binode}) implies that  $x^a = u^a$ so that the two valent node is not affected by relative locality.
This is fortunate because it means that the propagation of free particles is unaffected by relative locality. 

This is however not the case for the three and higher valent vertexes, as we are about to see.  Because of this, the implications of
relative locality appear in the interactions.

\subsection{The three valent interactions for emission and absorption of photons}

Now that we have warmed up with two valent nodes we are equipped to study the 
case of the interactions of atoms and photons involved in our model of gamma ray emission and
absorption.  

 In the model described above, we choose the 
 energy-momentum conservation laws for the emission vertices to be, 
\begin{eqnarray}
{\cal K}^1 &=&    (q^1 \ominus k^1  ) \ominus p^{1}  =0
\\
{\cal K}^2 &=&     (k^1  \ominus r^1  ) \ominus p^2   =0
\eea
These conservation law imply that the first photon momenta $p^{1}$ is equal to $q^{1}\ominus k^{1}$,
where $k^{1}$ is the momenta of the emitter after the photon emission.
Similarly  the second photon momenta $ p^{2}$ is equal to $ k^{1}\ominus r^{1}$, since now $k^{1}$ is the emitter momenta before the emission.

The conservation laws for the reception vertices are
\bea
{\cal K}^3 &=&   p^1  \oplus  ( \ominus k^2  \oplus q^{2} )  =0
\\
{\cal K}^4 &=& p^2  \oplus  ( \ominus r^2 \oplus k^2   )  =0
\end{eqnarray}
We have chosen a  specific  reception vertex  conjugate (in the sense that the addition is implemented in reverse order) to the emission vertex. 

We can directly compute the relations of relative locality associated with these conservation laws. We introduce a simple notion.
If $x_i^a (0)$ is the endpoint of a worline with momentum $k_a^i$ at a vertex where we impose the conservation law, $ {\cal K}_{\,\,b}=0$, we write
\f
 ({\cal W}_{x_{i}})_b^a  =  \pm \frac{\delta {\cal K}_b}{\delta k_a^i }
\ff
The sign is  choosen depending on whether the momenta are incoming $+$ or
outgoing $-$ to the vertex. 

We then can write
\bea
x_{i}&=& z_{i}{\cal W}_{x_{i}},\quad u_{i}= z_{i}{\cal W}_{u_{i}},\quad y_{i}=  z_{i}{\cal W}_{y_{i}}. 
\eea

We can  find the forms of the ${\cal W}$'s in terms of the basic, geometric parallel transport operators.
Lets us first exemplify this  calculation in great detail for $ {\cal W}_{x_{1}}$ say.
The momenta conjugated to $x_{1}$ is $q^{1}$. Thus, in order to do this calculation 
we first write that ${\cal K}^{1}= R_{\ominus p^{1}}R_{\ominus k^{1}}(q^{1})$ where $R_{p}(q)\equiv q\oplus p$ is the right multiplication. Then one take the differential of ${\cal K}^{1}$ with respect to $q^{1}$ and use the chain rule for derivation to 
get 
\be
 \rd_{q^{1}} {\cal K}^{1} = \rd_{q^{1}}(R_{\ominus p^{1}} R_{\ominus k^{1}})
 = (\rd_{(q^{1}\ominus k^{1})}R_{\ominus p^{1}})(\rd_{q^{1}}R_{\ominus k^{1}})
\ee
 We then use the definition (\ref{Rtransport}) of the right handed transport operator to evaluate this as 
 \be
 \rd_{q^{1}} {\cal K}^{1} =V_{((q^{1}\ominus k^{1})\ominus p^{1})}^{(q^{1}\ominus k^{1})}V_{(q^{1}\ominus k^{1})}^{q^{1}}
 \ee
Finally, we evaluate this expression at the value where ${\cal K}^{1}=0$, which implies that $ q^{1}\ominus k^{1} = p^{1}$
and we get the final expression:
\be
\rd_{q^{1}} {\cal K}^{1} = V_{0}^{p^{1}}V_{p^{1}}^{q^{1}}.
 \ee
  We can repeat this straightforward calculation to find the form of all the $\cal W$'s in terms of the basic, geometric parallel transport operators and we get:
 
\begin{eqnarray}
{\cal W}_{x_{1}} & = & \rd_{q^{1}} {\cal K}^{1} = V_{0}^{p^{1}}V_{p^{1}}^{q^{1}}
\\{\cal W}_{x_{2}} & = & \rd_{k^{1}} {\cal K}^{2} =V_{0}^{p^{2}}V_{p^{2}}^{k^{1}}
\\ {\cal W}_{x_{3}} & = & \rd_{q^{2}} {\cal K}^{3} =U_{0}^{\ominus p^{1}}U_{\ominus p^{1}}^{q^{2}}\\
 {\cal W}_{x_{4}} & = & \rd_{k^{2}} {\cal K}^{4} =U_{0}^{\ominus p^{2}}U_{\ominus p^{2}}^{k^{2}} 
\label{x-z}
\end{eqnarray}
for the incoming emitter/receptor particle momenta.
\begin{eqnarray}
{\cal W}_{u_{1}} & = & -\rd_{k^{1}} {\cal K}^{1} = -V_{0}^{p^{1}}U_{p^{1}}^{\ominus k^{1}} I^{k^{1}}
\\{\cal W}_{u_{2}} & = &- \rd_{r^{1}} {\cal K}^{2} =-V_{0}^{p^{2}}U_{p^{2}}^{\ominus r^{1}} I^{r^{1}}
\\ {\cal W}_{u_{3}} & = &- \rd_{k^{2}} {\cal K}^{3} =-U_{0}^{\ominus p^{1}}V_{\ominus p^{1}}^{\ominus k^{2}} I^{k^{2}}\\
 {\cal W}_{u_{4}} & = & -\rd_{r^{2}} {\cal K}^{4} =-U_{0}^{\ominus p^{2}}V_{\ominus p^{2}}^{\ominus r^{2}} I^{r^{2}}
\label{u-z}
\end{eqnarray}
for the outgoing emitter/receptor particle momenta.

The transport operator characterising the photon trajectories are given by
 \begin{eqnarray}
{\cal W}_{y_{1}} & = & -\rd_{p^{1}} {\cal K}^{1} = -U_{0}^{\ominus p^{1}}I^{\ominus p^{1}} = V^{p^{1}}_{0}
\\{\cal W}_{y_{2}} & = & -\rd_{p^{2}} {\cal K}^{2} = -U_{0}^{\ominus p^{2}}I^{\ominus p^{2}} = V^{p^{2}}_{0}
\\ {\cal W}_{y_{3}} & = & \rd_{p^{1}} {\cal K}^{3} = V_{0}^{ p^{1}} \\
 {\cal W}_{y_{4}} & = & \rd_{p^{1}} {\cal K}^{3} = V_{0}^{ p^{2}} 
 \end{eqnarray}
 
 In the first two lines we have used the bivalent vertex identity (\ref{binode}) to show that ${\cal W}_{y_{3}}={\cal W}_{y_{1}}$
 and ${\cal W}_{y_{4}}={\cal W}_{y_{2}}$.
 
\section{Deriving the time delay effect}

We have now developed a tool box that will enable us to compute expressions for the time delay we seek. 

\subsection{The basic strategy of the calculation}

The basic idea of the derivation that follows is to find an expression for the time delay, which represents the proper time between the events
when the detector registers the arrival of the two photons, which is also invariant under diffeomoprhisms of $\cal P$ and the transformations they induce
on the spacetime coordinates. 

To do this we derive a basic relation by following two routes from $z_1^a$ to $z_4^a$, forming a loop in spacetime.
First, let us recall how we derive the localisation equation in the usual case where  momentum space is flat and all interactions local in a single spacetime.
We start from the identity which is obviously true:
\f\label{Dsflat1}
 (z_4 -z_3 ) - (z_2-z_1 )  = (z_4 -z_2 )-(z_3 -z_1 )
\ff
Not to belabour the obvious, but this simple identity is invariant because all the $z_I^a$'s transform the same way because they all live in the
same linear space, which is the cotangent plane of the origin of $\cal P$.  

\subsection{Example of the strategy in special relativity}

In flat momentum space this can be evaluated easily as the parallel transport operators all reduce to the identity so
at each node locality prevails and
\f
z_I^a=x_I^a = y_I^a = u_I^a
\ff

We can use this plus (\ref{gammapt}) and (\ref{atompt})  to evalute (\ref{Dsflat1}) as
\be\label{Dsflat}
\hat{k}^{2}S_{2}- \hat{k}^{1}S_{1} = \hat{p}_{2} T_2- \hat{p}_{1}T_1 
\ee
where $\hat{p}=\hat{p}^{1}=\hat{p}^{2}$  is the direction of emission of the photon.

Assuming that the emitter and detector are at rest with each other, i-e that $\hat{k}^{1}=\hat{k}^{2}=\hat{k}$ 
we can show that (\ref{Dsflat}) implies that the two photons are emitted parallely (i-e $\hat{p}^{1}=\hat{p}^{2}$).
Taking the scalar product of (\ref{Dsflat}) with this common null direction allows us to
 conclude that, of course, $S_{2}=S_{1}$.
That is, in special relativity, the emission proper time delay is the same as the reception proper time delay.

Notice a remarkable feature of a simple calculation, which is that we started with a trivial identity in coordinates (\ref{Dsflat1}) and ended up with 
a somewhat less trivial identity of invariants, (\ref{Dsflat}).  The latter, which is a relation among physically observable proper times no longer depends on
coordinates because the physical relations are independent of the choice of the location of the origin of coordinates.  When locality becomes relative
locality, we will see that the relations between proper times are still independent of the choice of coordinate frame, at least to leading order, but the
relation between the proper times becomes dependent on the energies and momenta of the particles traveling on the worldlines.  

\subsection{First try at the time delay calculation}

In the general curved momentum case we want to apply the same strategy.
For the first try lets start from the same identity (\ref{Dsflat1}).

The difference is that we cannot evaluate  the expression directly in terms of the time of flight because of the distinction between the 
interaction coordinates $z_{i}$ and the particles cordinates $x_{i},u_{i}, y_{i}$.

We then need to develop a series of relations between the different coordinates.

We start with
\bea
z_{2}-z_{1} &=& (z_{2}-x_{2}) + (x_{2}-u_{1})+ (u_{1}-z_{1}) 
\nonumber \\
& =& z_{2}(1- {\cal W}_{x_{2}}) + \hat{k}^{1} S_{1} + z_{1}({\cal W}_{u_{1}}-1),
\eea
and similar expressions\footnote{From now on we work in Riemannian normal coordinates} for  other $z$'s differences.

Then
 (\ref{Dsflat1}) can be expressed as 
 \bea
 \hat{k}_2  S_2  -  \hat{k}_1  S_1 -  \hat{p}_{2} T_2+ \hat{p}_{1}T_1  = \Delta z
 \eea
 where
 \bea
  \Delta z  &=& z_{1}( {\cal W}_{u_{1}} -{\cal W}_{y_{1}})  
  +z_{2}({\cal W}_{y_{2}}-{\cal W}_{x_{2}}) 
  \nonumber \\ & &
  +z_{3}({\cal W}_{y_{3}}-{\cal W}_{u_{3}})+ z_{4}({\cal W}_{x_{4}}- {\cal W}_{y_{4}})
\eea
If one restrict to the case $\hat{k}^{1}=\hat{k}^{2}=\hat{k}$ where the emitter and detector are at rest with each other, taking the scalar product of the previous expression with $\hat{p}=\hat{p}^{1}=\hat{p}^{2}$ we get
\be
(S_{2}-S_{1}) = \Delta z \cdot \hat{p}
\ee
which shows that $S_{2}$ no longer  needs to be equal to $S_{1}$.

\subsection{Main derivation}

In order to get a definite prediction we need to be able to write the relations among 
the proper times, eliminating $z$'s from the identity. Similarly to what we did in  (\ref{Dsflat})  when momentum space was flat.
In order to do so we now follow a slightly different strategy:
We  start again from a basic identity that involves  following two routes from $z_1^a$ to $z_4^a$, forming a loop in spacetime.
We also use the fact that we would like this identity to be written in terms of quantities transforming covariantly under diffeomorphism and for that we  pull back any covector to the origin $T^{*}_{0}$ in such a way that we can eliminate the  intermediate coordinates $z_{I}$.

Lets start by noticing that 
\bea
 T_{1}\hat{p}_{1}{\cal W}_{y_{3}}^{-1} &=& (y_{3}-y_{1}){\cal W}_{y_{3}}^{-1} =  z_{3} - z_{1}{\cal W}_{y_{1}}{\cal W}_{y_{3}}^{-1} 
 \nonumber \\
 &=& z_{3}-z_{1}
\eea
and that we also have   
\be
 S_{2}\hat{k}_{2}{\cal W}_{u_{3}}^{-1} =(x_{4}-u_{3}){\cal W}_{u_{3}}^{-1} =  z_{4}{\cal W}_{x_{4}}{\cal W}_{u_{3}}^{-1} -  z_{3} 
\ee

Eliminating $z_{3}$ between the last two expressions we get an identity involving only $z_{4}$ and $z_{1}$
\be\label{z1}
z_{4}{\cal W}_{x_{4}}{\cal W}_{u_{3}}^{-1} -  z_{1}  =  S_{2}\hat{k}_{2}{\cal W}_{u_{3}}^{-1}+ T_{1}\hat{p}_{1}{\cal W}_{y_{3}}^{-1} 
\ee

Similarly, using 
\be
T_{2}\hat{p}_{2}{\cal W}_{y_{4}}^{-1}= z_{4}-z_{2},\qquad S_{1}\hat{k}_{1}{\cal W}_{x_{2}}^{-1} =  z_{2}- z_{1}{\cal W}_{u_{1}}{\cal W}_{x_{2}}^{-1}
\ee
and eliminating $z_{2}$ from these we get another identity only $z_{4}$ and $z_{1}$
\be\label{z2}
z_{4}- z_{1}{\cal W}_{u_{1}}{\cal W}_{x_{2}}^{-1} =  T_{2}\hat{p}_{2}{\cal W}_{y_{4}}^{-1} + S_{1}\hat{k}_{1}{\cal W}_{x_{2}}^{-1}
\ee

We can now eliminate $z_{4}$ between (\ref{z1}) and (\ref{z2}). This gives us the final expression that we are looking for
\bea  
{S_{2}K_{2}- S_{1}K_{1}
- T_{2}P_{2} +T_{1}P_{1} =  z_{1}\left({\cal W}_{u_{1}}{\cal W}_{x_{2}}^{-1} -{\cal W}_{u_{3}}{\cal W}_{x_{4}}^{-1}\right)} \label{maineq}
\eea
where we have introduced new momenta 
\bea\nonumber
K_{1} &\equiv&  \hat{k}_{1}{\cal W}_{x_{2}}^{-1}\\ \nonumber
 K_{2} &\equiv& \hat{k}_{2}{\cal W}_{x_{4}}^{-1} \\ \nonumber
P_{1} &\equiv& \hat{p}_{1}{\cal W}_{y_{4}}^{-1} \\
 P_{2}&\equiv& \hat{p}_{2}{\cal W}_{y_{3}}^{-1}{\cal W}_{u_{3}}{\cal W}_{x_{4}}^{-1} \label{Kk}
\eea

We now evaluate this expression assuming that all momenta are small
we approximate 
\be
(U_{p}^{q})_{a}^{b} \approx \delta_{a}^{b} - \Gamma_{a}^{cb}(p-q)_{c},\quad (V_{p}^{q})_{a}^{b} \approx \delta_{a}^{b} - \Gamma_{a}^{bc}(p-q)_{c}
\ee
a direct computation shows that 
at this order we have 
\bea
({\cal W}_{u_{1}}{\cal W}_{x_{2}}^{-1})_{a}^{b}& \approx&({\cal W}_{u_{3}}{\cal W}_{x_{4}}^{-1})_{a}^{b} \approx
\delta_{a}^{b} +  T_{a}^{bc}p^{1}_{c} 
\eea
so the RHS of (\ref{maineq}) vanish if one keep only first order terms in the non linear expansion.
The RHS is in fact proportional to a curvature, which will come into play at second to leading order.  
We postpone discussion 
of its implications to another paper.  

For the rest of this paper we will be interested in the implications of the leading order contributions to the
LHS of (\ref{maineq}). 

In order to evaluate the LHS at leading order we compute
\bea\nonumber
K_{1}^{b} &\approx&  \hat{k}_{1}^{b} - m_{1} \hat{k}_{1}^{a}\Gamma_{a}^{bc}\hat{k}^{1}_{c}\approx (\hat{k}_{1}V^{0}_{k_{1}})^{b},\\ \nonumber
 K_{2}^{b} &\approx& \hat{k}_{2}^{b} - m_{2} \hat{k}_{2}^{a}\Gamma_{a}^{cb}\hat{k}^{2}_{c}\approx (\hat{k}_{2}U^{0}_{k_{2}})^{b}\\ \nonumber
P_{1}^{b} &\approx& \hat{p}_{1}^{b} - E_{1} \hat{p}_{1}^{a}\Gamma_{a}^{cb}\hat{p}^{1}_{c}\approx (\hat{p}_{1}U^{0}_{p_{1}})^{b}\\
 P_{2}^{b}&\approx& \hat{p}_{2}^{b} - E_{2} \hat{p}_{2}^{a}\Gamma_{a}^{bc}\hat{p}^{2}_{c}\approx (\hat{p}_{2}V^{0}_{p_{2}})^{b} \label{Pp}
\eea
We see that, remarkably, to the approximation we are working, each of the new momenta  is a function only of the corresponding original momentum.  So $K_{1}$  is only a function of  $k_{1}$, and not of the others. This result is not obvious from the original definition (\ref{Kk}).

Thus, these momenta are, up to the approximation we are working with,  the original momenta, viewed as element of $T^{*}_{p}$,
 parallel transported to $T^{*}_{0}$, using either $U$ or $V$.
The final localisation equation is therefore similar to the naive one (\ref{Dsflat}), except that the momenta entering 
the equation are these parallel transported momenta. That is we have 
 \bea\label{Dsnonflat}
 {K}_2  S_2  -  K_1  S_1 =  P_{2} T_2- P_{1}T_1 .
 \eea
 In order to analyse the consequences of this equation we study separately the effect of non metricity and of torsion.
  
 \subsection{The case of non metricity}
 Let us first remark that since we work in Riemannian normal coordinates we can express the connection coefficients at the identity 
 entirely in terms of the torsion and non metricity tensors.
 Indeed if we denote by $\{{}_{\,\,c}^{ab} \}$ the Christofell symbol
 we have that 
 \bea\nonumber
 \Gamma^{(ab)}_{c} &=& \{{}_{\,\,c}^{ab} \} + \frac12g_{ci}\left(N^{abi}+N^{bai}- N^{iab} + T^{iab}+T^{iba}\right)\\
\Gamma^{[ab]}_{c} &=& \frac12  T^{ab}_{c} \label{TN}
\eea
 where we have defined
\be
N^{abc}\equiv \nabla^{a}g^{bc},\qquad T^{abc} =  T_{i}^{ab} g^{ic}.
\ee 
 In Rieman normal coordinates the Christoffel symbol vanish at the origin.
 This shows that the  connection is entirely determined by the non metricity tensor and the torsion tensor.
 Note that the antisymmetric part of the connection depends only on the torsion while the symmetric part depends 
 on both the torsion and the non metricity tensor.
 Note however that from the definition we have that in Rieman normal coordinates $2\Gamma^{a(b}_{i}g^{c)i}=N^{abc}$.
 
 We can now compute the norms  and scalar product of the newly defined vectors.
 Since these vectors are tangent vector at the origin (i-e $\in T_{0}$), we can use that the metric at the origin is simply the flat metric 
 $g^{ab}(0)=\eta^{ab}$ to compute norms and scalar product, we get:
 \be
 K_{i}^{2} \approx 1- m_{i} N^{\hat{k}^{i}\hat{k}^{i}\hat{k}^{i}}, \qquad  P_{i}^{2} \approx  - E_{i} N^{\hat{p}^{i}\hat{p}^{i}\hat{p}^{i}}
 \ee
 where we have defined $V^{\hat{k}}\equiv \hat{k}_{a} V^{a}$ and $E_{i}, i=1,2$ denotes the energy of the photons $i$.
 We see that the dispersion relation of the newly defined momenta are modified by the presence of the non metricity tensor.
 
In order to analyse (\ref{Dsnonflat}) we first assume that the  momenta $K_{1}$ and $K_{2}$ are paralell to each other.
When the torsion vanishes $U^{0}_{k}=V^{0}_{k}$ and if the emitter and detector mass are the same, which is what we now assume, this implies that $\hat{k}_{1} =\hat{k}_{2}$.
This express the fact that the emitter and detector are at rest with respect to each other.
We denote by 
$$\hat{K}_{a} = \frac{K_{a}^{1}}{|K^{1}|} = \frac{K_{a}^{2}}{|K^{2}|} $$ the common normalised direction.
The scalar product between $\hat{K}$ and $P_{i}$ will be needed, it gives: \be
 \hat{K}\cdot P_{i} = 1 - \frac{1}{2}\left( 
 E_{i}N^{\hat{k}{\hat{p}}_{i}\hat{p}_{i}} + m N^{\hat{p}_{i}{\hat{k}}\hat{k}} - m N^{\hat{k}\hat{k}\hat{k}}\right).
 \ee
If one denotes
\be
\Delta S \equiv |K_{2}| S_{2}- |K^{1}| S_{1}
\ee
 the identity (\ref{Dsnonflat}) becomes 
 \be\label{Dsfinal}
  \hat{K} \Delta S  = T_{2}P_{2}-T_{1}P_{1} 
 \ee
The photons' momenta can be decomposed as 
\be\label{PN}
P_{i}^{a} = (\hat{K}\cdot P_{i}) \hat{K}^{a} + \sqrt{(\hat{K}\cdot P^{i})^{2}-P_{i}^{2}} \,R^{a}_{i}
\ee
 where $R_{i}$ are unit spacelike vectors perpendicular to $\hat{K}$.
 The equation (\ref{Dsfinal}) imposes three conditions. First we have that 
 \be\label{ds1}
 \Delta S = (\hat{K}\cdot P_{2}) T_{2}-(\hat{K}\cdot P_{1}) T_{1}
 \ee
 together with 
 $R_{1}^{a}= R_{2}^{a}$ and  finally 
 \be{\label{dt1}}
 T_{2} \sqrt{(\hat{K}\cdot P_{2})^{2}-P_{2}^{2}} - T_{1} \sqrt{(\hat{K}\cdot P_{1})^{2}-P_{1}^{2}} =0.
 \ee
 taking the difference of (\ref{ds1}) and (\ref{dt1}) we obtain 
  the final expression for the time delay
 \bea\label{Timedelay}
  \Delta S = \{(\hat{K}\cdot P_{2})- \sqrt{(\hat{K}\cdot P_{2})^{2}-P_{2}^{2}}\} T_{2}
  \nonumber \\
  -\{(\hat{K}\cdot P_{1})  -  \sqrt{(\hat{K}\cdot P_{1})^{2}-P_{1}^{2}}\} T_{1}
 \eea
 This expression clearly shows that if  $P_{i}$ are null vectors, which happens when the non-metricity vanishes, 
 there is no time delay.
 
 We can finally use our approximation to compute the time delay to first order in $\Gamma$ in order to arrive at the simpler expression
 \bea
  \Delta S &\approx& \frac{P_{2}^{2}}{2\hat{K}\cdot P_{2}} T_{2}-\frac{P_{1}^{2}}{2\hat{K}\cdot P_{1}} T_{1}
  \nonumber \\
  &\approx& -\frac12 (T_{2} E_{2}N^{\hat{p}_{2}{\hat{p}}_{2}\hat{p}_{2}} - T_{1} E_{1}N^{\hat{p}_{1}{\hat{p}}_{1}\hat{p}_{1}})
 \eea
 In the limit where the first   photon is of very low energy and the second one  is of very high energy, we get the 
 time delay
 \be
 \boxed{ \Delta S  =- \frac12 T_{2}E_{2} N^{+++}}
 \ee 
 which involves the components of the non metricity tensor along the photon direction $\hat{p}_{2}\approx \hat{p}_{1}$ denoted $+$.
 
 This means that even if the high and low energy photon are emitted at the same time ( in the frame of reference of an observer local to the emission) they are not detected as arriving at the same time!
Remarkably this time delay is proportional to the time of propagation between emission and detection, an effect growing with the distance of observation.
Another noticeable feature of the result is that the masses of the emitter and receptor atoms do not enter the final result,
so this result is independent of the nature of the emission and reception.

\subsection{The case of  non-vanishing torsion}

In this subsection we now assume that the  torsion does not vanish, while we assume that the symmetric part of the connection vanishes.
According to (\ref{TN}) this imply that the non metricity is fixed by the torsion tensor.  Similar results will be obtain if we chose the non metricity itself to vanish;  we choose to illustrate this choice for its simplicity.
In this case both $P_{i}$ are null vectors (at first order in $\Gamma$) and from (\ref{Timedelay}) we conclude that
there is no time delay $\Delta S =0$.

However, there is still a non trivial effect due to torsion:
From (\ref{PN}) we can see that $P_{1}$ and $P_{2}$ are proportional to each other and we call this common 
null direction $e^{+}$ (in the following we denote by $e_{-}$ the covector $e_{-}^{b}= e^{+}_{a}\eta^{ab}$).
But in the presence of torsion this doesn't mean that  the direction of  photon propagation  
$\hat{p}_{1}, \hat{p}_{2}$ are equal to each other.
Indeed, under the hypothesis that the connection is purely anti-symmetric, inverting the relations (\ref{Pp})  we can express
\bea\label{pe+}
p_{1}^{a} &\approx& (1+a_{1}) e_{-}^{a} + \frac12 E_{1} T_{-}^{+ a}\\
p_{2}^{a} &\approx& (1+ a_{2}) e_{-}^{a} - \frac12 E_{2} T_{-}^{+ a}
\eea
where $a_{i} = O(\Gamma)$.
This shows that $p_{2}$ is obtained from $p_{1}$ by an infinitesimal rotation in the plane determined by $ e_{a}^{+}$ and the vector $ T_{-}^{+ a}\equiv e_{-}^{c} e^{+}_{b}T_{c}^{ba}$.
More precisely,  the notion of a rotation between two null directions is defined with respect to an timelike observer.
In our case we consider an observer at rest with respect to $\hat{k}= e_{0}$ and we want to compute  what is 
the rotation angle $\Delta \theta$ between the null directions $p_{1}$ and $p_{2}$ that such an observer experiences.

In order to do so,  let $e^{+}= e^{0}+e^{1}$ where $e^{0}=\hat{K}$ is the timelike frame of emission and absorption.
Let us  denote by $B_{01}$ the infinitesimal boost in the direction of $e^1$ and $R_{1A}$, $A=2,3$ the infinitesimal rotation
in the  plane spanned by $e_1$ and $ e_A$. This means that $B_{01}^{ab} = 2\delta_{0}^{[a}\delta_{1}^{b]}$  and $R_{1A}^{ab}= 2\delta_{1}^{[a}\delta_{A}^{b]}$. (\ref{pe+}) implies that  the null momenta $p_{i}$ are boosted  in the direction of the timelike frame and rotated away from the common direction $e_{1}$, that is
\be\label{rp}
p_{i}\approx e^{\alpha_{i} B_{01} + \theta_{i} \hat{T}^{A} R_{1A} } e_{-} , 
\ee
where the rotation angles are  given by
\be\nonumber
\theta_{1} = \frac12 E_{1} |T_{-}^{+}|,\quad \theta_{2}=- \frac12 E_{2} |T_{-}^{+}|,\quad  |T_{-}^{+}|\equiv \sqrt{|T_{-}^{+a}\eta_{ab}T_{-}^{+b}|} . 
\ee
and the direction of rotation $\hat{T}$ is a unit vector  given by 
\be
T_{-}^{+A} = |T_{-}^{+}| \hat{T}^{A}
\ee

This implies that the rotation angle between $p_{1}$ and $p_{2}$ is given  by
\be\label{Dtheta}
\boxed{\Delta \theta = \frac{E_{1}+E_{2}}{2}  |T_{-}^{+}| }
\ee
This result shows that in the presence of torsion even if the two photons which are emitted in parallel and at the same time 
are detected at the same time, they appear to be coming from two different directions in the sky.
 The difference is proportional to the average photon energy.  We may call this effect {\it dual gravitational lensing.}

\section{Conclusions}

The special theory of relativity taught us to give up the notion that there exists an absolute space, by giving a sensible description to phenomena like length contraction that would be paradoxical if we kept that notion. The compensation is a new notion of absolute spacetime, within which this and other phenomena have a consistent description.  Analogously, the framework of relative locality introduced in \cite{PRL} encompasses phenomena that are apparently paradoxical if one maintains the construct of an absolute spacetime, and gives them a consistent description in terms of an absolute phase space.  This phase space, being a cotangent bundle of a curved momentum space, has no section corresponding to a universal spacetime, as all the sections are momentum and energy dependent.

The apparent paradoxes were pointed out in \cite{Unruh,sabine}, which greatly served to stimulate this work.  In this paper we have seen how the framework of relative locality cleanly resolves the apparent paradoxes of locality in theories with curved momentum space, leading to new kinds of observable phenomena. AT first order we have established two types of phenomena:
First, there is a differential time delay in the arrival time of simultaneously produced photons, relative to an observer at the emission events, which measure non-metricity of momentum space. Second there is a dual gravitational lensing effect which provides a measure of  torsion is momentum space. 
 A discovery of either of these phenomena would be an indication we really do not live in a spacetime, but live in a phase space.

In order to establish these results we have  studied the Einstein localisation procedure in the context of relative locality
associated with the process of emission and reception of two photons of different energies.
We have seen that at first order in $E/m_{P}$ the relationship between momenta and proper time expressing 
the localisation procedure is the same as in the usual special relativity case.
The main difference is that the momenta entering the localisation formula are no longer the same than the 
particle momenta determining the spacetime velocity. One is obtained as the parallel transport of the other using the momentum space connection. When this connection is non metric this implies the the photon momenta entering the localisation formula 
possess an energy dependent mass that results in an effective time delay.
When the connection is  torsionfull the particle momenta is twisted with respect to the momenta entering the localisation formula, resulting in dual gravitational lensing effect.

One of the things we learn from the study of this example is that the question of whether the speed of light is dependent on energy or not, in theories with curved momentum space or non-commutative coordinates-is coordinate dependent and hence unphysical.  To ask a physical question one must construct a model of a series of physical events defined by invariants, such as we have done here.  The delay time we have computed here is a physical proper time, but where the origin of the delay lies-whether in an energy dependence of the speed of light or in non-localities at the emission or detection events, is a coordinate dependent question.  

The two effects we have found here may be surprising because they do not conform to the expectation that quantum gravity effects take place at short distances, at a scale set by the Planck length, $l_p =\sqrt{\hbar G}$.  Instead, the time delay effect is of the form $\delta x \approx x (\mbox{Energies}/m_p) $ and so is relevant when very large distances are involved.  And, remarkably, dual gravitational
lensing, predicts an effect at an angle $\Delta \theta \approx \mbox{Energies}/m_p$ which is independent of the distance the light travels.  These remind us of ultraviolet/infrared mixing, a phenomena associated with non-commutative geometry-and indeed the interaction coordinates $z^a$ are non-commutative, as shown in  \cite{PRL}.  More generally, they represent a regime of quantum gravitational phenomena characterized by $m_p$ which may be detectible even when $l_p$ can be ignored.

We should stress some limitations of the present analysis.  First, in our analysis we have set $\hbar=0$. There are likely additional quantum effects which could affect the interpretation of the gamma ray burst time delays, such as a stochastic element due to anomalous spreading of wave packets \cite{GioLee,sabine,mespread}.     Second, the model we have used is very much simplified in several ways.  The particles are modeled as interacting relativistic particles.  This we believe is sufficient to describe the relevant physics in the approximation we use here in which $\hbar=G_{Newton}=0$, but it will be important to extend the description of relative locality to field theory and quantum field theory.  Furthermore, we model the emitter and detector as each single atoms-that can be improved.  

Third, our analysis is only correct to leading order in ${1}/{m_p}$.  There will be additional effects of order ${1}/{m_p^2}$ given by components of the curvature
tensor on momentum space.  We postpone consideration of these effects to another paper.  We note however, that assuming the usual Planck scale the effects of
order ${1}/{m_p^2}$ are not relevant for current experiments.  They however might be relevant for accelerator experiments on the assumption of a $Tev$ Planck scale as is necessary in the framework of large extra dimensions.  

Also our analysis relies on a specific choice of trivalent vertices. We would like to establish to what extend the results described here depends on the specific choice of vertex interaction. We hope to come back to this important question in a future publication.

A last limitation is that our analysis is entirely phenomenological.  Whatever the quantum theory of gravity is, if it gives well defined predictions for physical phenomena,  it must make predictions for the classical, non-gravitational regime described by here and in \cite{PRL}.  We therefore propose that all aspirants to a quantum theory of gravity be challenged to produce predictions for the geometry of momentum space which can be subject to experimental test.  We note that this requirement is satisfied by quantum gravity in $2+1$ dimensions, where a specific non-trivial geometry of momentum space has been shown to emerge in the low energy limit \cite{3d}).  

Within these limitations we can however assert that the current experimental limits from observations of time delays from the Fermi satellite\cite{fermi} do serve to bound
the components of the non-metricity tensor of momentum space.  The bound published in \cite{fermi} can be interpreted as implying,
\f
|N^{+++}| \lesssim \frac{1}{0.6 m_p}
\ff
This means that the framework of relative locality is already constrained to order of the Planck scale by existing experiments.  This justifies the assertion in the title of this paper and gives us reason to hope that future experiments will give us more information about the geometry of momentum space.

\section*{Acknowledgements}

We are first of all very grateful to our fellow relative-localists Giovanni Amelino-Camelia and Jerzy Kowalski-Glikman for comments, criticism and encouragement every step of the way along this work.  We are also very grateful to Nima Arkady-Hamid, Dan Gottesman, Sabine Hossenfelder for critical and helpful comments.  Our presentation of this work was improved by the opportunity to present it in talks at Columbia University, Perimeter Institute and ILQGS.   Research at
Perimeter Institute for Theoretical Physics is supported in part by
the Government of Canada through NSERC and by the Province of
Ontario through MRI.

\end{document}